\documentclass[12pt]{iopart}
\usepackage{iopams}

\usepackage[nonamebreak,numbers,square,comma,sort&compress]{natbib}

\providecommand{\newblock}{\ }
\usepackage{color}
\newlength{\imwidth}
 \definecolor{ForestGreen}{rgb}{0.13333333,  0.54509804,  0.13333333}

\usepackage{graphicx}
\usepackage[colorlinks]{hyperref}

\begin{document}
% \ShowCorrect

\title{Factorisation of laser pulse ionisation probabilities in the multiphotonic regime}
\author{R. Della Picca, J. Fiol  and P. D. Fainstein}
\address{
CONICET and Centro At\'{o}mico Bariloche, Comisi\'{o}n Nacional de Energ\'{\i}a At\'{o}mica, Avda E. Bustillo 9500, 8400 Bariloche, Argentina}
\eads{renata@cab.cnea.gov.ar}

\date{\today}

\begin{abstract}
We present a detailed study of the ionisation probability of H and  H$_{2}^{+}$ induced by a short intense laser pulse. Starting from a Coulomb-Volkov description of the process we derive a multipole-like expansion where each term is factored into two contributions: one that accounts for the effect of the electromagnetic field on the free-electron final-state and a second factor that depends only on the target structure. Such a separation may be valuable to solve complex atomic or molecular systems as well as to interpret the dynamics of the process in simpler terms. We show that the series expansion converges rapidly, and thus the inclusion of the first few terms are sufficient to produce accurate results. 
\end{abstract}

\maketitle
%%%%%%%%%%%%%%%%%%%%%%%%%%%%%%%%%%%%%%%%%%%%%%%%
\section{Introduction}

The interaction of strong and short laser pulses with atoms and molecules has received renewed attention in the recent past, both experimentally and theoretically \cite{Brabec2000RMPp545,Brabec2008_SFL,Milosev2006JPBpR203,Becker2005JPBpR1,Joachai2000AIAMOP_HLPp225}, mainly because the  advances in laser technology have made feasible time-resolved measurements of atomic and molecular processes.
These advances made possible new experimental investigations of atomic and molecular processes on an ultrashort time-scale, and under ultra intense laser radiation. These techniques lead the way to measurement of highly nonlinear phenomena, and even their control is now possible \cite{Krausz2009RMPp163,Corkum2007NPp381,Hentsch2001Np509}.

On the theoretical side, since the first works by Kulander \cite{Kulande1987PRAp2726,Kulande1987PRAp445}, more than two decades ago, many computational techniques have been developed to solve the three-dimensional time-dependent Schr\"{o}dinger equation (TDSE) for single-electron systems \cite{Choi2002PRAp25401,Bauer2006CPCp396,Grumgrz2010PRAp43408,Zhou2011PRAp33406}. The advances in computing power allows us to perform these computations in a question of minutes nowadays. However, the computation power demands imposed by precision numerical solutions of the Schr\"{o}dinger equation quickly scales out-of-reach when the complexity of the systems increases. 

Alternative, complementary approaches based on perturbative expansions could provide results at relatively low-computer cost at the expense of some precision-loss. 
Traditionally, the Keldysh-Faisal-Reiss or Strong-Field Approximation (SFA) \cite{Keldysh1964ZETFp1945,Faisal1973JPBpL89,Reiss1980PRAp1786} (see also \cite{Milosev2006JPBpR203}) has been employed to theoretically describe these processes when the field is intense.
On the contrary, when the field is weak, a First Order Perturbation theory (FPA) is sufficient to describe the process \cite{Schiff1968_QM}.
In the intermediate regime, where the interaction of the electrons with the nucleus and with the electromagnetic (EM) field are of similar strength, none of the potentials may be neglected. To describe ionisation processes in this regime was developed the Coulomb-Volkov approximation (frequently called CV2$^-$ or CVA) \cite{Duchate2000JPBpL571,Duchate2001PRAp53411,Guichard2007PSp397,Duchate2002PRAp23412}, where both interactions with the ejected electron are taken into account at the same level in the final-state.

The CVA approach has been widely used to investigate the ionisation of atomic hydrogen \cite{Duchate2002PRAp23412,Arbo2008PRAp13401}, alkali metal atoms \cite{Duchate2001PRAp13405}, simple molecules \cite{Rodrigu2005JPBp2775} and positronium \cite{Rodrigu2006NIMBp105}. Furthermore, there are several implementations of the Coulomb Volkov approximation, that may be used in a wide range of conditions: the renormalised CV (RCV2$^-$) extends the CVA to the non-perturbative regime \cite{Gayet2005JPBp3905}, the modified CV (MCV2$^-$) and close-coupling CV (CC-CV2$^-$) 
introduces the coupling to intermediate bound states \cite{Rodrigu2004PRAp53402,Rodrigu2011JPBp5603}, and the doubly distorted CV (DDCV) \cite{Arbo2010TEPJDp193,Graviel2012JPBp5601} includes the distortion by the laser field also in the initial state.

Alternatively, Dimitrovski \textit{et. al.} \cite{Dimitro2004PRLp83003,Dimitro2005PRAp43411} have presented analytical formulas for ionisation by very short pulses, that are independent of the laser intensity, in the context of First Magnus Approximation (FMA).  When weak and short field are applied, FPA and FMA can be used to obtain the sudden approximation \cite{Dimitro2004PRLp83003,Schiff1968_QM} where the ionisation amplitude is proportional to the momentum transfer and the dipole transition matrix.

The interaction with intense laser fields could induce, besides multiphoton ionisation (MPI), another important phenomenon: High order Harmonic Generation (HHG). In a semiclassical recollision picture it proceeds in three steps: first the electron is released from the target, then the electron interacts with the laser field, and in a third step, if the electron returns to the atom, radiative recombination may take place.
Alternatively, in the third step, an elastic scattering or an (e,2e) process may  take place, giving place to High-order Above-Threshold Ionisation (HATI) or Non-Sequential Double Ionisation (NSDI), respectively \cite{Joachai2000AIAMOP_HLPp225, Lin2010JPBp122001}. 

Despite the well-known shortcomings of the Strong Field Approximation (SFA), such as the discrepancies for calculations on different gauges, three-steps processes may be understood within its simple framework. The first-order SFA is able to describe `direct' ionisation, but a second order term is necessary to include the post-ionisation interaction between the free electron and the target. In contrast, this mechanism is already included in the first-order CVA, providing both direct and rescattering amplitudes \cite{Milosev2006JPBpR203}.
Recently, an alternative approach to this problem, called Quantitative Rescattering theory (QRS), was developed by Lin \textit{et. al.} \cite{Lin2010JPBp122001}. In this approach the yields for HHG, HATI and
NSDI can be expressed as the product of the returning electron wave packet probability with photo-recombination, elastic electron scattering, and electron-impact ionisation cross sections, respectively.
Also recently, in a full quantum mechanical description, Frolov \textit{et. al.} have deduced an analytical factorisation of the spectra in terms of an electron wavepacket and the cross-section of photo-recombination (for HHG), and elastic electron scattering (for HATI), see \cite{Frolov2011PRAp21405,Frolov2011PRAp43416,Frolov2012PRLp213002} and references therein.

This type of yield factorisation allows the study of more complex atoms or molecules, by separating the roles of the laser-pulse and the parent-ion, and allowing to accurately extract information about the target, (see for example \cite{Lin2010JPBp122001, Le2008JPBp081002,Wang2012PRAp15401,Chen2009PRAp33409}) offering a promising tool for dynamic chemical imaging with temporal resolution of few femtoseconds.

In this work we study the above threshold ionisation (ATI) of atomic hydrogen by short laser pulses within the CVA approach.  We derive an expansion of the transition matrix in powers of the laser-field vector potential, whose first term is sufficient to describe ionisation in the multiphotonic regime. This first-order term is written as the product of an one-photon ionisation transition-matrix and an integral factor depending on the laser pulse.

We compare the ATI spectra with TDSE results and analyse the convergence of the expansion, how each factor of this approximation contributes to the electronic spectra, and present some useful applications of this factorisation.
Finally, we also investigate the ionisation from some excited states of the hydrogen atom, and from $\mathrm{H}_{2}^{+}$ employing exact wavefunctions for both the initial and final channels. 

Atomic units are employed, except where otherwise stated.

%%%%%%%%%%%%%%%%%%%%%%%%%%%%%%%%%%%%%%%%%%%%%%%%
\section{Review of the Coulomb-Volkov approximation}

Let us consider the ionisation of an atomic or molecular system by interaction with a finite laser pulse of duration $\tau$. In the length gauge, the action of the pulse may be described as a time-dependent force produced by the electric field. Thus, the time-dependent Schr\"{o}dinger equation (TDSE) reads
\begin{equation}\label{Q:tdse-1}
\rmi\, \frac{\partial \Psi( \bi{r},t)}{\partial t} 
= \left[ \frac{\hat{\bi{p}}^{2}}{2} + V(\bi{r},t) + \bi{F}(t)\cdot\bi{r} \right]  \Psi(\bi{r},t) \,.
\end{equation}

We consider a finite laser pulse, that exerts a force on the system given by its electric field
\begin{equation} \label{Q:electric-field-Ft-0}
\bi{F}(t)= F_{0}\, \sin{(\omega (t-t_{0}))}\; \sin^{2}{(\pi\,t/\tau)} \; \hat{ \varepsilon} 
\end{equation}
for $0 < t < \tau $, and that vanishes at all other times. The pulse is characterized by its central frequency $\omega $, the pulse duration $\tau$, the polarisation vector $\hat{\varepsilon}$, and the phase-shift, chosen as $\omega t_{0} = (\omega \tau - \pi) /2$ for symmetric pulses.

While the perturbation vanishes outside the time-interval $\Delta T= (0, \tau)$ we can write the ionisation transition-matrix, in its prior form, as
\begin{equation} \label{Q:tif} 
  T_{fi}^{-}= -\rmi \int_{0}^{\tau} \big\langle \Psi^{-}_{f}(t) \big| \bi{F}(t)\cdot \bi{r} \big| \phi_{i}(t) \big\rangle   \, \rmd t \,.
\end{equation}
Here $\Psi^{-}_{f}(t)$ is the exact wavefunction for the final state with ingoing boundary conditions, and $\phi_{i}(t)$ is the asymptotic target wavefunction in absence of external fields.

The double differential ionisation probability in energy and angle of the emitted electron is obtained from the transition matrix magnitude as
\begin{equation}\label{difP}
 \frac{d P_ {fi} }{dE d\Omega}= k \left| T_{fi}\right|^{2}\, .
 \end{equation}

The Coulomb-Volkov approximation (CVA) is obtained by replacing the exact final wavefunction $|\Psi^{-}_{f}(\bi{r},t)\rangle$ by a product of factors corresponding to the solution of two separated problems: one for the isolated field-free atom and one containing the effects of the electron evolving in the external electromagnetic (EM) field. In order to get the correct asymptotic behaviour the plane-wave part must be corrected \cite{Duchate2000JPBpL571,Duchate2001PRAp53411}. This procedure is similar to the one carried-out in continuum-distorted-wave (CDW) theories developed for ion-atom collisions many years ago \cite{Dewanga1992CAMPp317,Dewanga1994PRp59,Fainste1991JPBp3091}. In this approximation the final-state wavefunction takes the form \cite{Duchate2000JPBpL571,Duchate2001PRAp53411}
\numparts
\begin{eqnarray}
  \label{Q:defin-coul-volk-state}
  \fl    \chi^{-}_{f}(\bi{r}, \bi{k}, t) = \frac{\rme^{\rmi \bi{k}\cdot \bi{r}}}{(2 \pi)^{3/2}}\, D^{-}_{C}(\bi{k}, \bi{r}) \, D^{-}_{\bi{A}}(\bi{k}, \bi{r}, t) \, \rme^{-\rmi E_{f} t} \\
  \fl    D^{-}_{C}(\bi{k}, \bi{r}) = (2 \pi)^{3/2}\,\rme^{-\rmi \bi{k}\cdot \bi{r}}\,\varphi^{-}_{f}(\bi{r}) \label{Q:def-DC-CV}   \\
  \fl    \label{Q:def-Da-CV} D^{-}_{\bi{A}(t)}(\bi{k}, \bi{r}, t) = \rme^{\rmi \bi{A}(t) \cdot \bi{r}}\, \exp{\left[- \frac{\rmi}{m}\, \bi{k} \cdot \int_{\tau}^{t}  \bi{A}(t')\, \rmd t'  \right]}\nonumber \\
  \times \exp{\left[ - \frac{\rmi}{2}\int_{\tau}^{t}  \left( A(t') \right)^{2}\, \rmd t'  \right]} \\
  \fl    \bi{A}(t) = - \int_{\tau}^{t} \bi{F}(t')\, \rmd t'
\end{eqnarray}
\endnumparts
Here $\varphi^{-}_{f}(\bi{r})$ is the final-state of the electron ejected with momentum $\bi{k}$, corresponding to energy $E_{f}=k^{2}/2$, and with ingoing wave boundary conditions. We have explicitly set $m=1$ and $Z=-1$ for the electron mass and charge, respectively. 
 For a pure Coulomb potential interaction (hydrogen atom) the distortion factor takes the familiar form:
\[
  D^{-}_{C}(\bi{k}, \bi{r}) = N^{-}(\nu) \,{_1F_1}\left(  \rmi \nu;1; - \rmi (k r + \bi{k} \cdot\bi{r} ) \right) \, ,
\]
where the normalization factor is defined by
\(N^{-}(\nu) = \Gamma(1 - \rmi\nu) \rme^{-\pi \nu/2} \), and \(\nu=-1/k\) is the Sommerfeld parameter. For general, more complex, targets the distortion factor must be obtained numerically, and is defined in terms of the final-state target eigenfunction $\varphi_{f}$ as given by (\ref{Q:def-DC-CV}).

After some manipulation, the CVA transition matrix may be written as \cite{Duchate2002PRAp23412,Guichard2007PSp397}
\numparts
  \begin{eqnarray}
    T_{fi}^{CV}&= f(0)g(0) +  \int_{0}^{\tau} \rmd t\, h(t) \, f(t)\,g(t)  \label{Q:Tif-CVA} \\
 \fl \mathrm{where} & \nonumber \label{Q:where1}\\
h(t)&= \rmi\,\left( \omega_{fi} +  \bi{k}\cdot\bi{A}(t) +  {\bi{A}}^{2}(t)/2 \right) \, ,\label{Q:Tfi-CVA-ht}\\
f(t)&= \exp{( \int_{\tau}^{t} h(t')\,\rmd t' + \rmi \omega_{fi} \tau} )\, ,\label{Q:Tfi-CVA-ft}\\
g(t)&= \int d\bi{r}\, \varphi_{f}^{-*}(\bi{r})\, \exp{(-\rmi \bi{A}(t)\cdot \bi{r})}\, \varphi_{i}(\bi{r}) \, , \label{Q:Tfi-CVA-gt}
  \end{eqnarray}
\endnumparts
and $\omega_{fi}= E_{f}-E_{i}$. Since $A(\tau)=0$, the additional term involving $g(\tau)$ in the above expression vanishes due to the orthogonality of the initial and final wavefunctions.

%%%%%%%%%%%%%%%%%%%%%%%%%%%%%%%%%%%%%%%%%%%%%%%%
\subsection{Applicability of CVA} \label{ApplyCVA}

The Coulomb-Volkov approach (CVA) is obtained by replacing the exact final-state wavefunction in the transition matrix (\ref{Q:tif}) by the well-know Coulomb-Volkov wavefunction (CVF) \cite{Jain1978PRAp538,Cavaliere1980JPBp4495}. Many studies have been performed to delimit the domain in wich CVF may be used. In particular, Kornev and Zon used an algorithm based in the time-dependent Siegert theorem to determinate the accuracy of CVF  \cite{Kornev2002JPBp2451}. They found that the CVF is applicable if the laser amplitude and frequency hold the condition $F_0 < 0.1 \,\omega$. However, strictly speaking, as the authors noted, this  requirement is not a sufficient condition to determine the accuracy of CVF.  

On the other hand, the CVA is a time-dependent distorted-wave theory that
has been successfully employed to describe several processes in the interaction with laser fields \cite{Duchate2000JPBpL571,Duchate2001PRAp53411,Guichard2007PSp397,Duchate2002PRAp23412,Arbo2008PRAp13401,Duchate2001PRAp13405, Rodrigu2005JPBp2775, Rodrigu2006NIMBp105}.

The authors  of these previous works pointed-out that CVA gives accurate electron energy distributions when:
(i) the ionisation probability is small, or in other words, the population of the initial state remains basically unchanged during the interaction (perturbative conditions),  and (ii) the photon energy is greater than the ionisation potential $\omega > I_{p}$ (transitions to intermediate states before ionisation are energetically prohibitive). Apart from this, very recently Gravielle \textit{et. al.} \cite{Graviel2012JPBp5601} have analyzed the distortion in the initial state due to the laser field, they concluded that it should be taken account when the quiver amplitude, characterized by the parameter $\alpha_0=F_0/\omega^2$, is comparable with the mean radius of the initial electronic distribution $r_0$.

So, summarizing some previous studies, the CVA may be applied under the following conditions:
\begin{enumerate}\renewcommand{\theenumi}{\alph{enumi}}
\item $F_0/\omega< 0.1$ for accurate CV wavefunctions,
\item $P_ {fi} \ll 1$ for perturbative conditions,
\item $ I_{p} \le \omega$ no transitions to excited bound states,
\item $F_0/\omega^2 \ll r_0 $ initial displacement negligible. 
\end{enumerate}

Recently, several modifications were introduced to extend the applicability of CVA, beyond the above conditions. In order to overcome the restrictions on (b) the renormalised CV (RCV2$^-$) includes the evolution of the initial state together with the interaction \cite{Gayet2005JPBp3905}.  Modified CV (MCV2$^-$) and the  close-coupling CV (CC-CV2$^-$) approximations, that include the transitions from the initial to intermediate states, may be used for laser photon energies smaller than the target ionisation energy (c) \cite{Rodrigu2004PRAp53402,Rodrigu2011JPBp5603}. Finally, restriction (d) may be solved by including the distortion of the laser both in the initial and final states, as has been done in the doubly distorted CV (DDCV) \cite{Arbo2010TEPJDp193,Graviel2012JPBp5601}. In these works good agreement with numerically-solved TDSE results was found, even outside the very restrictive condition (a).

%%%%%%%%%%%%%%%%%%%%%%%%%%%%%%%%%%%%%%%%%%%
\section{Multipolar expansion of the T-matrix}

It has been shown that the probability may be evaluated efficiently in the CVA approximation, rendering good results with relatively low computational cost in the case of atomic hydrogen. However, its extension to more complex systems such as many-electron atoms or even small molecular ions can considerably increase computation times. It is desirable to develop approximations that keep the good performance of the CVA but may be applied to laser-induced ionisation of large molecules.

In the range of applicability of the CVA theory, the amplitude of the field $F_{0}$ is small and 
in most cases the magnitude of the vector field $\bi{A}(t)$ is also small. In that case we may expand the exponential in $g(t)$ in \eref{Q:Tfi-CVA-gt} or alternatively in $D^{-}_{\bi{A}(t)}$ (Eq.~\ref{Q:def-Da-CV}), and keep only the first, most significant, terms
\begin{equation} \label{expansion}
  \exp{(-\rmi \bi{A}\cdot \bi{r})}= 1  - \rmi \bi{A}\cdot \bi{r} - \frac{1}{2} (\bi{A} \cdot \bi{r})^{2} + \dots
\end{equation}

We obtain a series expansion for the transition matrix $T_{fi}^{CV}= T^{(0)} + T^{(1)} + T^{(2)} + \dots$, whose terms are given by

\numparts
  \begin{eqnarray}
    \label{Q:Tif-CVA-serie}
    T^{(0)} &=   \langle \varphi_{f}^{-}| \varphi_{i} \rangle \\
    T^{(1)} &= - \rmi\, \langle \varphi_{f}^{-}| \bi{r} | \varphi_{i} \rangle  \cdot  \left[ f(0)\bi{A}(0) +\int_0^\tau \rmd t\, f(t)\,h(t) \bi{A}(t)  \right] \\
    T^{(n)} &= \frac{(-\rmi)^n}{n!} \,L^{(n)} \, M^{(n)}
  \end{eqnarray}
\endnumparts
where we have defined for $n\ge 2$:
\begin{eqnarray}
L^{(n)}&=& \langle \varphi_{f}^{-}|(\hat{\varepsilon}\cdot \bi{r})^{n} | \varphi_{i} \rangle    \\
M^{(n)}&=&  f(0) A^n(0) +\int_0^\tau \rmd t\, f(t)\,h(t) A^n(t) \\
       &=& n \, \int_0^\tau \rmd t\, f(t)\,F(t) (A(t))^{n-1}  \nonumber
\end{eqnarray}

Each term of this expansion is written as the product of two factors, from very different origin: $M^{(n)}$ accounts for the laser perturbation and its effect on the final state while $L^{(n)}$ is sensitive to the target structure. 
 
Here we have explicitly used the linear polarization of the laser field $\bi{A}(t)=\hat{\varepsilon}\, A(t)$ to separate the spatial and temporal integrals for higher-order terms. Note that $ T^{(1)}$ is factorisable 
independently of the polarization of the laser.

To our knowledge, this type of expansion has not been performed before as it is presented here. However very recently, the expansion \eref{expansion} has been carried out  in the context of laser pulse propagation in dielectric materials \cite{Bourgeade2012PREp56403}. 
Also, Guichard \textit{et. al.} \cite{Guichard2007PSp397} have expanded the exponential of the Volkov distortion 
$D^{-}_{\bi{A}}$, but the detailed study of the expansion was outside the scope of their work.
Furthermore, the first-order has been previously employed in 
two-color XUV+IR photoionisation  \cite{Kazansky2010PRA82p33420}  
and in our previous work of laser-induced ionisation of atoms and molecules \cite{DellaPicca2012ICPEAC}.

%----------------------------------------------------------------
\subsection{DipA}
%----------------------------------------------------------------

If we keep only the first non-vanishing order of the above expansion, the density of probability \eref{difP} is approximated by
\begin{equation} \label{TmatDipA}
  T^{(1)} = - \rmi \; \bi{L}^{(1)} \cdot \bi{M}^{(1)}\,,
\end{equation}
allowing us to decouple the transition matrix as the product of two factors, where $ \bi{L}^{(1)}= \langle  \varphi_{f}^{-}|  \bi{r}  | \varphi_{i} \rangle $ is the one-photon ionisation transition matrix, and
\begin{equation}  \label{MFf}
  \bi{M}^{(1)} = \int_{0}^{\tau} \bi{F}(t) \, f(t)\,\rmd t \,.
\end{equation}

The above approximation relies on the condition $\bi{A}(t) \cdot \bi{r} \ll 1$, for that reason we call it \textit{Dipole Approximation} or DipA. While the amplitude of the vector potential $\bi{A}(t)$ is proportional to
the electric field amplitude $F_{0}$ and inversely proportional to the frequency $\omega$, the range of validity of the approximation is constrained by these two parameters. In principle some of these constrains are
already considered in Coulomb-Volkov approximations, as elaborated in section \ref{ApplyCVA}. 

However, for a given EM field DipA approximation may fail for highly-excited states, because the radius of the atomic or molecular system increases.
We note that this approximation may also be obtained by neglecting completely  the exponential $\exp{(\rmi \bi{A}(t) \cdot \bi{r})}$ in the Volkov state \eref{Q:def-Da-CV}. This factor contains the spatial dependence, mixed with the vector potential EM field. Thus, by neglecting it, space and time result completely decoupled in the transition matrix integrals~\eref{Q:tif}.

The Keldysh parameter $ \Gamma \equiv \sqrt{(I_p/2U_p)} \sim 1$ defines the limit between multiphotonic and tunneling regimes for ionisation processes. Here the ionisation potential energy is $I_{p}$ and the ponderomotive energy $U_{p}=F_{0}^2/4\omega^2$.  Then we expect that DipA will work well in the multiphotonic regime, when $ \Gamma \gg 1$, since in this case $A \sim 1/ \Gamma$ will be small.

To analyse the spectra obtained in DipA we study the different factors in (\ref{TmatDipA}). As we mentioned before, $ \bi{L}^{(1)} $ is the ionisation transition matrix due to the absorption of one fictitious photon with frequency $\omega_{fi}= E_{f}-E_{i}$. For eigenfunctions of the unperturbed Hamiltonian, the matrix element can be written equivalently either in the length or velocity gauge:
\begin{equation} \label{Tphoto}
  \bi{L}^{(1)} =   \langle  \varphi_{f}^{-}|  \bi{r}  | \varphi_{i} \rangle = -   \frac{1}{\omega_{fi}} \langle  \varphi_{f}^{-}| \nabla    | \varphi_{i} \rangle
\end{equation}
 This term does not contain any information on the EM field, and will be the same for all laser pulses.
On the other hand, since the function $f$ can be factorised as $f(t) = j(t) \, \exp{(\rmi \omega_{fi} t )}$, the contribution of the field can be written as
\begin{equation}
  \bi{M}^{(1)} =  \int_{0}^{\tau} \rmd t\,   \bi{F}(t) \,j(t) \, \rme^{\rmi \omega_{fi} t }
\label{Mpulse}
\end{equation}
Here $j(t)$ is the part of the Volkov state that is taken into account in DipA 
\begin{eqnarray}
  j(t) = &   \exp{\left[ \rmi\, \bi{k} \cdot \int_{\tau}^{t}  \bi{A}(t')\, \rmd t'   \right]}\nonumber \\  &\times \exp{\left[ \frac{\rmi}{2}\int_{\tau}^{t}  \left( A(t') \right)^{2}\, \rmd t'  \right]}
\end{eqnarray}

\Eref{Mpulse} has the aspect of a Fourier transform. However, the function $j$ and the \textit{Fourier frequency} $\omega_{fi}$ are not independent, since both depends on the electron energy $E_f$.
For sufficiently low intensity of the field, $j \sim 1$ and this equation defines exactly the Fourier transform of the EM pulse, resulting in the first Born approximation \cite{Guichard2007PSp397}. 
Also, it is important to note that the factor $\bi{M}^{(1)}$ depends on the target only through the binding energy $E_i$. For this reason, spectra for different targets, or initial states with different binding energies can be reproduced fairly accurately, by only shifting the kinetic electron energy.

In the following sections we analyse the multipolar expansion, and calculate explicitly the first and second orders. CVA spectra for H atoms and $\mathrm{H}_{2}^{+}$ molecular ions are compared with TDSE results and the spectra obtained employing the first- and second-order transition matrix in the multipolar expansion of CVA. In order to avoid introducing additional sources of errors we employ exact wavefunctions for both the initial and final states.

%%%%%%%%%%%%%%%%%%%%%%%%%%%%%%%%%%%%%%%%%%%
\section{ Spectra for H}

%--------------------------------------------------------------
\subsection{Comparison of different theories}

In this section we compare CVA and DipA spectra to the results obtained by numerically solving the TDSE with the code QPROP \cite{Bauer2006CPCp396}. For CVA calculations we employed equations \eref{difP} and \eref{Q:Tif-CVA}, while in the DipA case we considered \eref{difP} and \eref{TmatDipA}.

\begin{figure}
\centering
\includegraphics[angle=0,width=\imwidth]{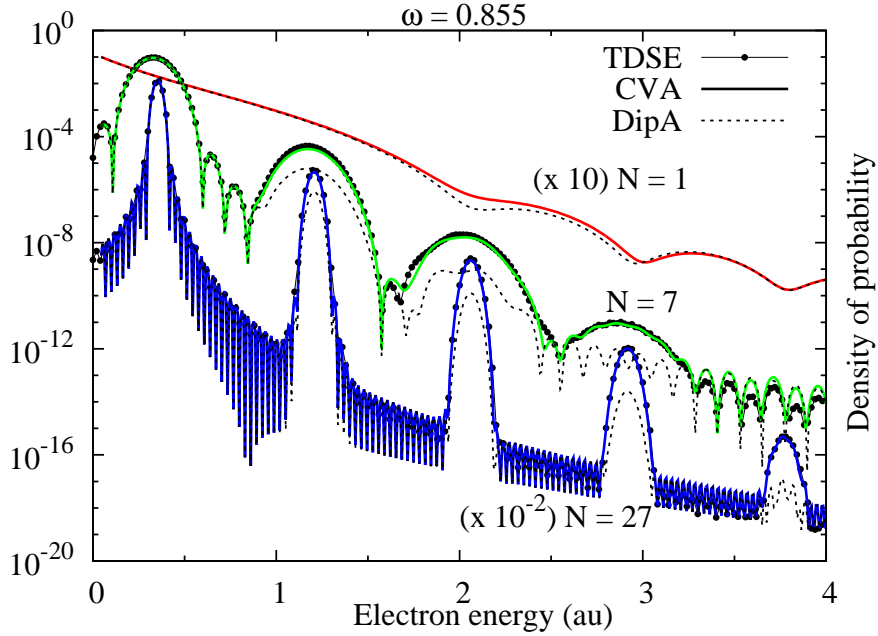}
\caption[]{(colour online) Electron spectra for ionisation of H($1s$) as function of the ejected electron energy for a laser pulse with N = 1 (up), N = 7 (middle) and N = 27 (bottom) cycles. Laser frequency $\omega=  0.855$~a.u., $F_0=0.05$~a.u. ($\Gamma = 17.1$). Full line: CVA, dotted line: Dipole Approximation (DipA)  and full line with circles: TDSE results.}\label{F05w085Atom}
\end{figure}

\begin{figure}
\centering
\includegraphics[width=\imwidth]{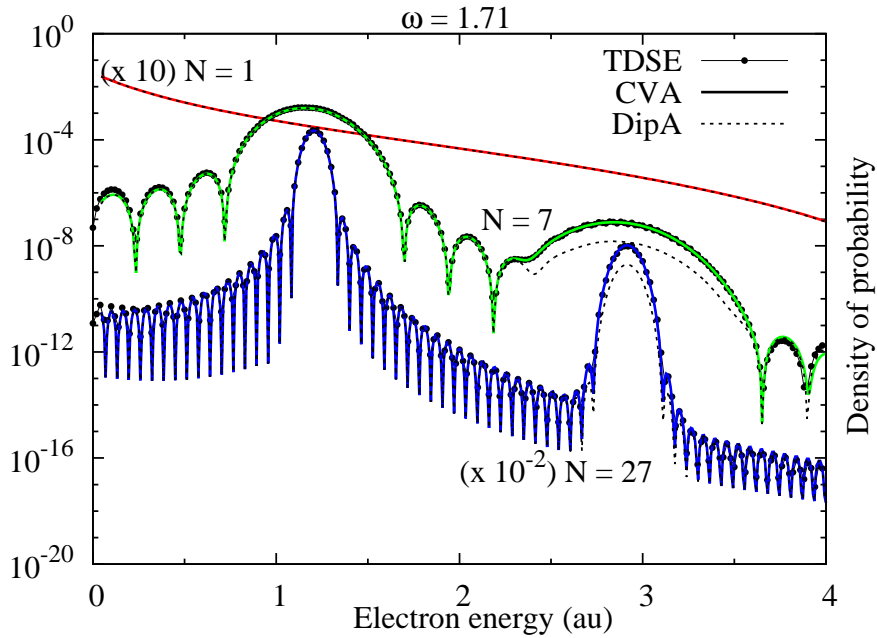}
\caption[]{(colour online) Idem Fig. \ref{F05w085Atom} with  $\omega= $ 1.71a.u. ($\Gamma$ = 34.2)}
\label{F05w171Atom}
\end{figure}

\begin{figure}
\centering
\includegraphics[width=\imwidth]{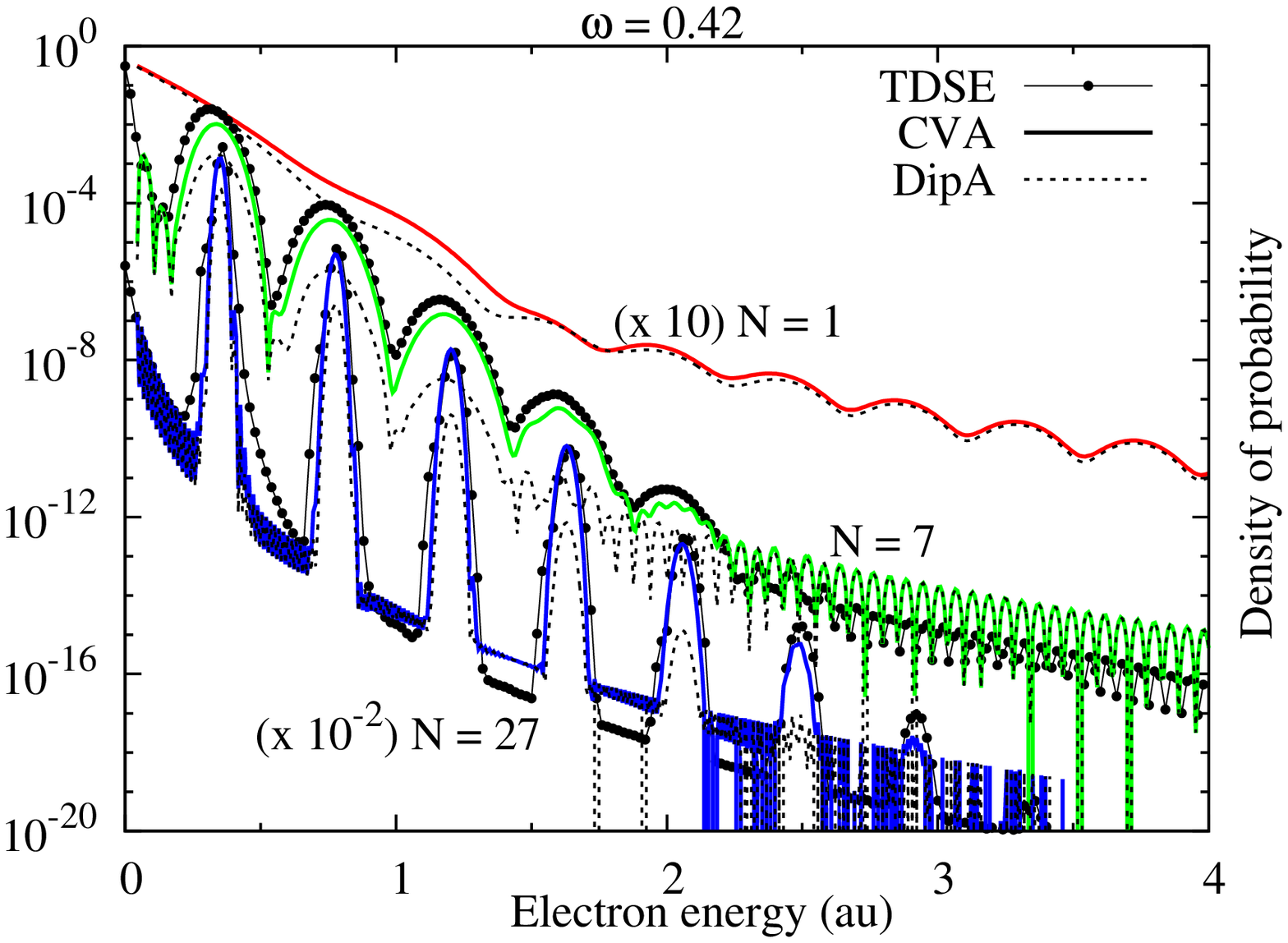}
\caption[]{(colour online) Idem Fig. \ref{F05w085Atom} with  $\omega= $ 0.427a.u. ($\Gamma$ = 8.55)}
\label{F05w042Atom}
\end{figure}

In figures \ref{F05w085Atom}, \ref{F05w171Atom} and \ref{F05w042Atom}  we present the density of probability (DOS) for the ionisation of atomic hydrogen ($1s$). The spectra, as function of the electron energy, were obtained integrating over all electron-emission directions.
We consider laser pulses of intensity  $F_0= 0.05$~a.u., duration corresponding to N = 1, 7 and 27 cycles,  and frequencies $\omega= $ 0.855  (Fig.~\ref{F05w085Atom}), 1.71 (Fig.~\ref{F05w171Atom}), and 0.427~a.u. (Fig.~\ref{F05w042Atom}). The results for N = 1 and 27 are multiplied by 10 and 10$^{-2}$, respectively, for better visualisation. 

In general, we observe in these figures a good agreement between CVA, TDSE and DipA results, specially for high frequencies of the laser. These examples are within the limits of applicability of the CVA approximation; thus they are almost indistinguishable from the numerically exact results given labelled TDSE.
At lower frequencies the disagreement between CVA and TDSE has been analysed by Duchateau \textit{et. al.~}\cite{Duchate2002PRAp23412,Guichard2007PSp397}, and they pointed out that the CVA fails because it does not consider transitions to intermediate excited states. This mechanism is more important when the ionisation energy is larger than the laser frequency.
We note that this case is in the limit of validity of the CV wavefunction (see point (a) of Section \ref{ApplyCVA}).
 
Moreover, we observe the formation of above-threshold-ionisation (ATI) peaks when increasing the number of cycles of the pulse. 
These structures correspond to the absorption of $n$ photons at the value energy $E_n = E_i +n \omega -U_p$ (see for example \cite{Milosev2006JPBpR203}).

%--------------------------------------------------------------
\subsection{Analysis of DipA }
\label{S:disc-la-aprox}

\begin{figure}[!hbtp]
  \centering
  \includegraphics[width=\imwidth]{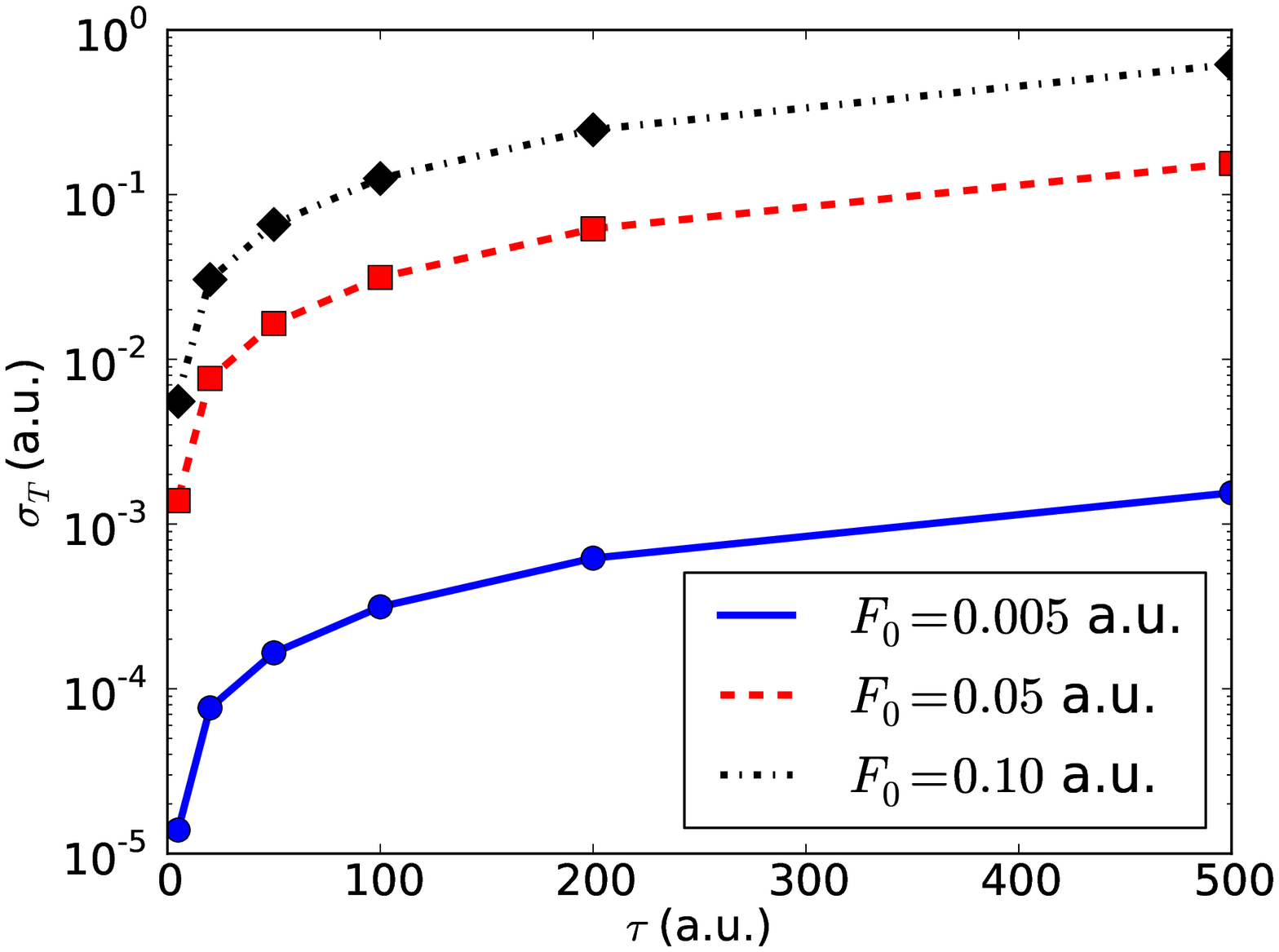}
  \caption{Comparison of hydrogen ($1s$) total ionisation probabilities by laser pulses  of frequency $\omega= 0.855$~a.u, for several intensities. Lines correspond to full CVA calculations, while symbols were obtained in the first order DipA.}
  \label{F:comparison-xs-totales}
\end{figure}

Let us now discuss the DipA results, plotted with dotted line in figures \ref{F05w085Atom} to \ref{F05w042Atom}.
In these figures we observe that the shape of the spectra is well reproduced by the dipole approximation, but there are disagreements in the high-order ATI peaks, corresponding to multiple-photon absorption, where DipA underestimates the ionisation probabilities. In all cases, the first peak, that is the most important in magnitude, is well reproduced by the DipA. As a consequence, total ionisation probabilities are accurately given by this approximation.
This is observed in \fref{F:comparison-xs-totales}, where we show total ionisation probabilities for a frequency $\omega= 0.855$~a.u, as function of the laser pulse duration, and for several EM intensities. We can observe that the first-order approximation (DipA) reproduces very well the full calculations in a very extended range of laser pulse parameters. 

As we mentioned before, the intensities of the secondary ATI peaks are not well described by DipA. This means that in this regions there are couplings between time and space, i.e: the exponential factor $\exp{(-\rmi \bi{A}\cdot \bi{r})}$ plays an important role in the description of the ionisation spectra.

The $\omega = 0.427$~a.u.~case of~\fref{F05w042Atom} is different since the ATI peaks are closer to each other and shifted to the left.  The first peak would appear at negative energies, below the ionisation threshold. Thus, the first peak corresponds to the absorption of two photons, worsening appreciably the comparison. This is an expected result, according with the above description, for low frequencies or high amplitudes $A$.

We note that for large electron energies, outside the ATI peaks, the background of the spectra is also well reproduced.  This fact can be explained noting that for high energies the contribution of the term $\bi{k} \cdot \int   \bi{A}(t)\, dt $ is more important than the contribution of $\bi{r} \cdot  \bi{A} $. Then, this last term can be omitted in the exponential of the Volkov state~\eref{Q:def-Da-CV}, resulting precisely in DipA.

Surprisingly, short pulses corresponding to $N = 1$ cycle are well reproduced by DipA independently of the laser frequency (figures \ref{F05w085Atom} to \ref{F05w042Atom}).
For very short pulses and in the context of the First Magnus Approximation (FMA), Dimitrovski \textit{et al.} \cite{Dimitro2004PRLp83003,Dimitro2005PRAp43411} found that the ionisation amplitude is independent of the pulse shape and, therefore, all pulses are equivalent if they have the same value of $ \bi{A}(0)$. 
In particular, the results are the same that would be obtained for a flat pulse, constant in time.
In connection with the present Dipole Approximation, we note that pulse amplitudes $\bi{A}(t)$ that vary slowly in time, allow naturally the decoupling of spatial and temporal integrals in the transition matrix.

Furthermore, in the cases of short pulses and weak fields, FMA and first order perturbation theory (FPA) can be combined into the \textit{sudden approximation} \cite{Schiff1968_QM,Dimitro2004PRLp83003}, where the ionisation amplitude is described by \eref{TmatDipA}, but simplifying the factor $f(t)=1$ in \eref{MFf}.
Our proposed DipA presents the same level of computation ease that the sudden approximation, but being valid for an extended range of laser parameters, as we can observe in \fref{F:comparison-xs-totales}.

%--------------------------------------------------------------
\subsection{Applications of DipA}
\label{S:apli-la-aprox}

DipA enable us to understand the features of the spectra, allowing us to separate those that arise from the pulse parameters from those that are exclusively due to the nature of the target. 

For example, because we consider ionisation of H atoms from the ground state in the cases presented in figures~\ref{F05w085Atom} to \ref{F:comparison-xs-totales}, in all cases the contribution of $\bi{L}^{(1)}$ is the same.  Its squared modulus, integrated in electron emission direction, is proportional to the photoionisation cross-section  and is a monotonically decreasing function of the energy (see for example eq.~37 of \cite{Dellapi2011PRAp33405}).  

On the contrary, the factor $\bi{M}^{(1)}$, is different for each of the nine spectra plotted in \fref{F05w085Atom} to \ref{F05w042Atom}, corresponding to the nine different laser pulses. 
For symmetric pulses with $N = 1$ cycle, the shape of the pulse is basically a peak, then the Fourier transform predicts a widely spread spectra, as observed from the above figures.
Increasing the number of cycles, the Fourier transform of the pulse narrows considerably, as a delta-like function, evaluated in the central frequency of the pulse. This behavior gives rise to the first ATI peak (one-photon absorption of central frequency $\omega$).
In order to analyse the formation of the secundary ATI peaks we can expand the exponential in $j(t)$ in powers of the argument. Each of these terms contributes  to an harmonic of the original pulse (multiphoton absorption), for this reason the Fourier transform shows the ATI peaks centered in positions that are multiples of $\omega$. 
Also, when increasing the intensity of the pulse, more terms in the exponential expansion are needed, explaining the formation of new ATI peaks as the intensity increases (see for example figure 1 of \cite{Duchate2002PRAp23412}).  
 
Summarising, we can note that the general shape of the spectra is defined  by $\bi{M}^{(1)}$: the number of ATI peaks, their amplitude and width could be estimated from the analysis of this factor, as well as the dependence on the carrier envelope phase (CEP), fixed to $\omega \,t_{0}$ in this work.   

One important application of this approximations is in the calculation of DOS for complex systems, like large molecules where the time integration of target form factors as in (\Eref{Q:Tif-CVA}) would be prohibitively time-consuming. In the DipA it is possible to construct the spectra for laser pulse ionisation by multiplying the one-photon ionisation transition-matrix of this complex system by the factor $\bi{M}^{(1)}$  with the information of the pulse. This approximation is valid for sufficiently high-frequency pulses, as those generated from typical HHG spectra \cite{Paul2001Sp1689,Bartels2002Sp376,Sansone2006Sp443,Mashiko2008PRLp103906}.
Conversely, it is possible ot obtain information about target structure from  the emission spectra by compare the experimental data with a decomposition as the one obtained in DipA model.
This type of experiments, where extreme-ultraviolet attosecond-light pulses are used as source to emit electrons and obtain ``tomographic images'' of the targets have only recently been achieved \cite{Hentsch2001Np509,Kienber2004Np817,Billaud2012JPBp4013}.

%--------------------------------------------------------------
\subsection{Ionisation from excited states and $2^{\mathrm{nd}}$-order approximation}

\begin{figure}
\centering
\includegraphics[angle=0, width=\imwidth]{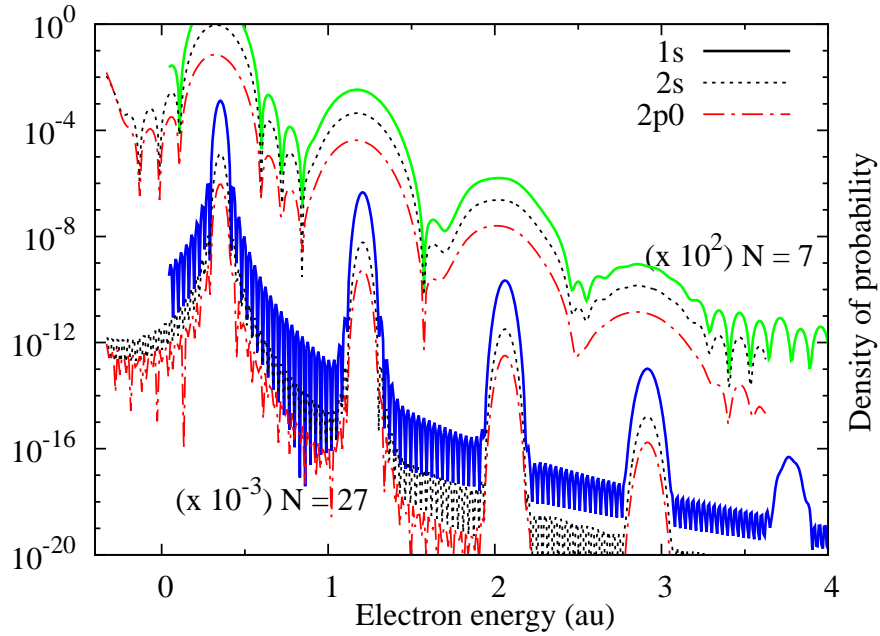}
\caption[]{(colour online) Ionisation spectra for H from $1s$ solid line, $2s$ dotted line and $2p_0$ dashed dotted line for laser pulses with $\omega=$ 0.855 a.u., $F_0=$ 0.05 a.u., N = 7 and 27 cycles. The ionisation spectra from $2s$ and $2p_0$ are shifted to lower energy in 0.375 a.u. }
\label{Excited}
\end{figure}

In this section we want to analyse the spectra for ionisation of hydrogen atoms from excited states.  In ~\fref{Excited} we present CVA ionisation probabilities for H atoms from the $1s$, $2s$ and $2p_0$ states. For the excited states we shifted the spectra in energy considering the difference in their binding energies, $E_1 - E_2 = -0.375$~a.u. As a result, the energy position of the ATI peaks for the different spectra are matched.  We observe that the shape of the spectra are very similar in all cases, but the ionisation probabilities are lower for excited states. These two facts can be understood in the DipA context: as we discussed before, the effect of the target on $\bi{M}^{(1)}$ produces a shift in the spectra that we have just corrected. On the other hand, the intensity of the spectra is modulated by the
different one-photon transition-matrix $\bi{L}^{(1)}$ in each case.

\begin{figure}
\centering
\includegraphics[angle=0, width=\imwidth]{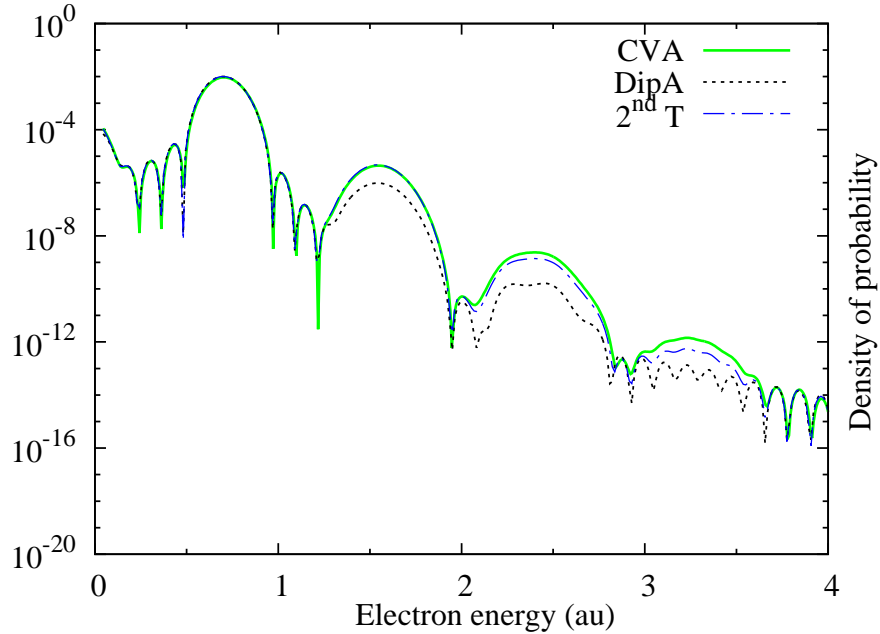}
\caption[]{(colour online) Ionisation spectra for H from $2s$ for laser pulses with $\omega=$ 0.855 a.u., $F_0=$ 0.05 a.u. and N = 7 cycles. Full line (green): CVA, dotted line (black): DipA and dashed dotted line (blue): $2^{\mathrm{nd}}$ order calculations.    }
\label{Excited2}
\end{figure}

In \fref{Excited2} we show also, the ionisation probabilities from H($2s$), comparing the CVA results with DipA. Like in the $1s$ case (see \fref{F05w085Atom}), DipA underestimates the magnitude of ATI peaks for multiple photon absorption.  Evidently, the first non-vanishing term in the multipolar expansion of the T-matrix is not enough to describe the ionisation process. In order to achieve a better representation we include a second term, i.e.~$T \approx T^{(1)} + T^{(2)}$  (see \ref{Q:Tif-CVA-serie}) in the differential ionisation probability \eref{difP}.
This calculation is labelled as second-order ($\mathrm{2^{nd}\,T}$) in \fref{Excited2}. We observe that the addition of the second term is sufficient to reproduce accurately the spectrum in an extended energy range. Visible differences appear only at high electron energies, at the third and fourth ATI peaks, where still shows a dramatic improvement over the first order.

%%%%%%%%%%%%%%%%%%%%%%%%%%%%%%%%%%%%%%%%%%%
\section{Spectra for H$_{2}^{+}$}

In this section we study the ionisation of fixed-in-space H$_{2}^{+}$ molecular ions by short laser-pulses.  The nuclei of the molecule, having charges $Z_a = Z_b = 1$, are fixed at the internuclear distance $R=2$~a.u. Employing spheroidal coordinates and using standard computational methods we calculate exactly the initial bound $\varphi_{i}(\bi{r})$ \cite{Hadinge1989JPBp697}, and final continuum $\varphi_{f}^{-}(\bi{r})$ \cite{Ponomar1976JCPp183,Rankin1979JCPp437} states of the molecule.  Also the plane-wave $\exp{(-i \bi{A}(t)\cdot \bi{r})}$ is expressed as a sum of spheroidal harmonics \cite{Morse1953_MOT,Dellapi2011PRAp33405}.  With these wave functions the factor $g(t)$ in (\ref{Q:Tfi-CVA-gt}) is evaluated numerically for the CVA calculation.  To obtain DipA spectra, $\bi{L}^{(1)}$ is
computed in the same way that in previous works \cite{Dellapi2007PRAp032710,Dellapi2009PRAp32702,Dellapi2011PRAp33405}.

The use of accurate initial and final wavefunctions introduces a heavy computational cost in the numerical evaluation of the factor $g(t)$ as a sum of partial terms. This is a consequence of the partial-wave descomposition of $\varphi_{f}^{-}$ and $\exp{(-\rmi\bi{A}\cdot \bi{r} ) }$ (full calculation).  The simplifications introduced by the use of the first-order DipA improve the computations times by a factor of approximately 300.
Besides those large speed-up factors obtained in the first-order calculations, we observed that accurate results are produced by including up to order 3 in the  multipolar expansion of the T-matrix. In this case the speed factor, \textit{i.e.} time ratio between full and  $\sum_{\le 3} T^{(n)}$ calculation is about 50.

\begin{figure}[!hbtp]
\centering
\includegraphics[angle=0, width=\imwidth]{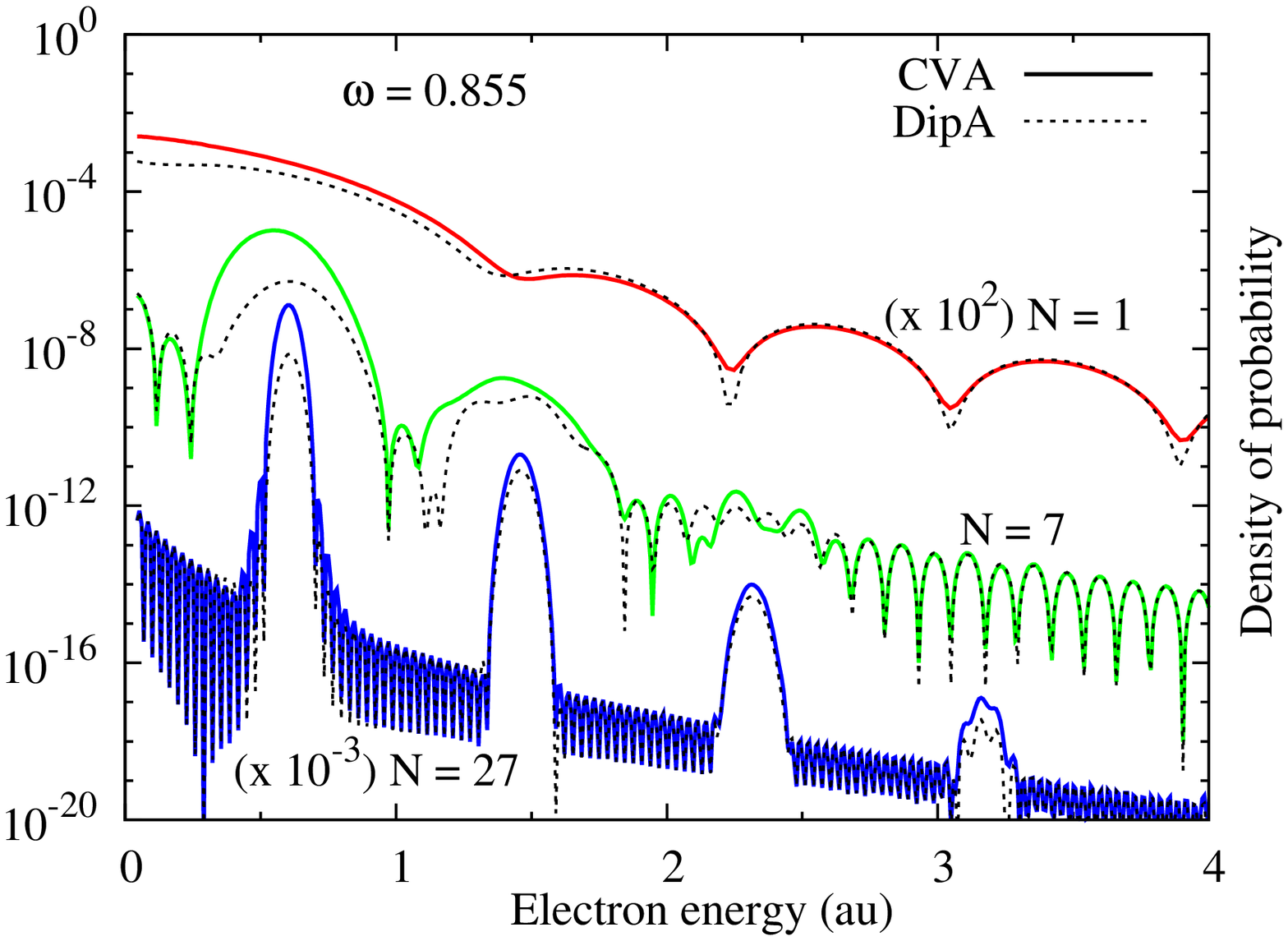}
\includegraphics[angle=0, width=\imwidth]{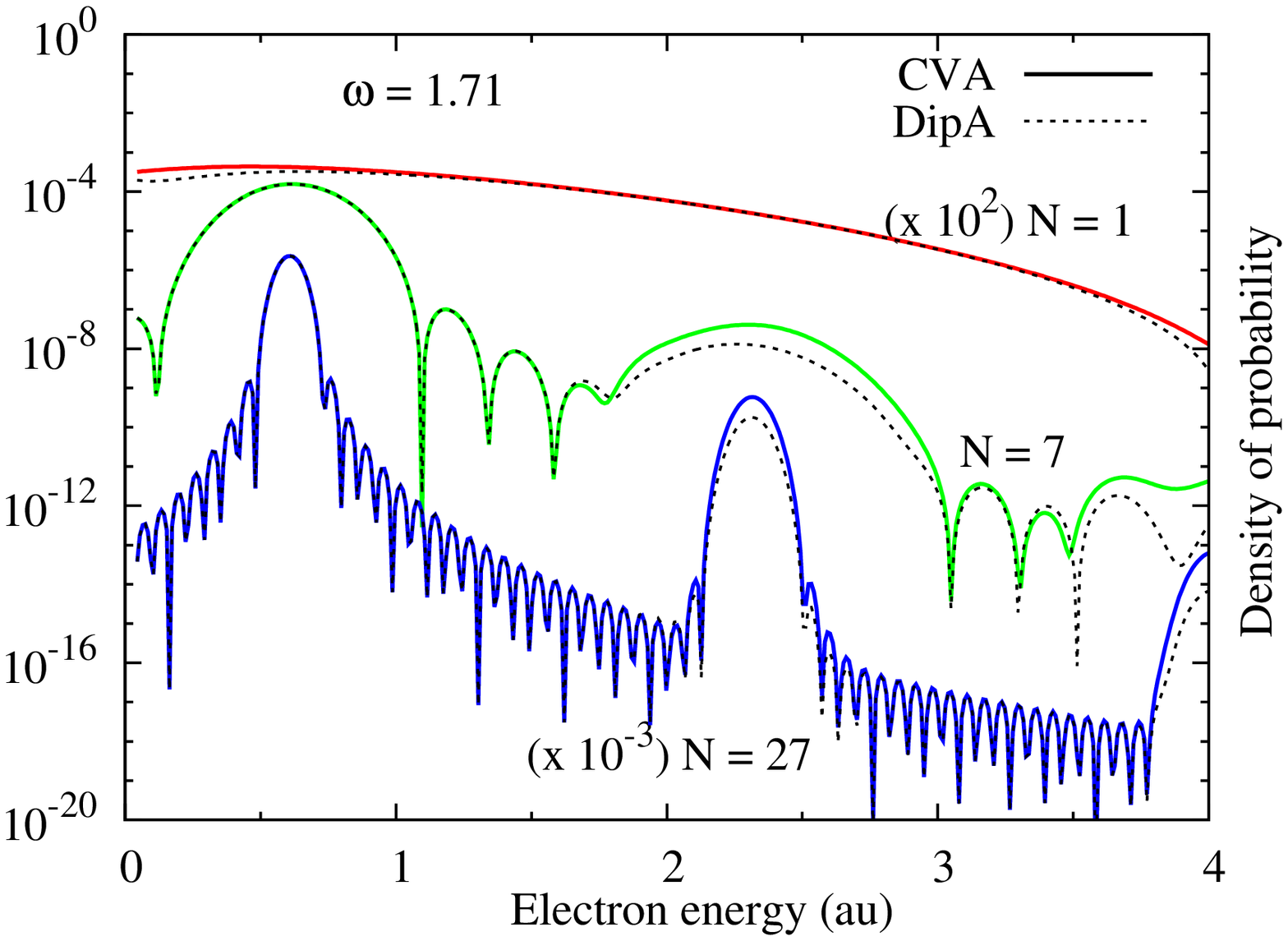}
\caption[]{(colour online)  Electronic spectra for ionisation of H$_{2}^{+}$($1s\sigma$) as function of the electron energy, for emission in the forward direction. We have considered laser-pulses of $N = 1, 7$, and $27$ cycles, with frequencies $\omega=  0.855$ and $1.71$~a.u., and an amplitude $F_0= 0.05$~a.u. Full line: CVA and dotted line: DipA.}
\label{F05w085y171H2}
\end{figure}

\begin{figure}
\centering
\includegraphics[angle=0, width=\imwidth]{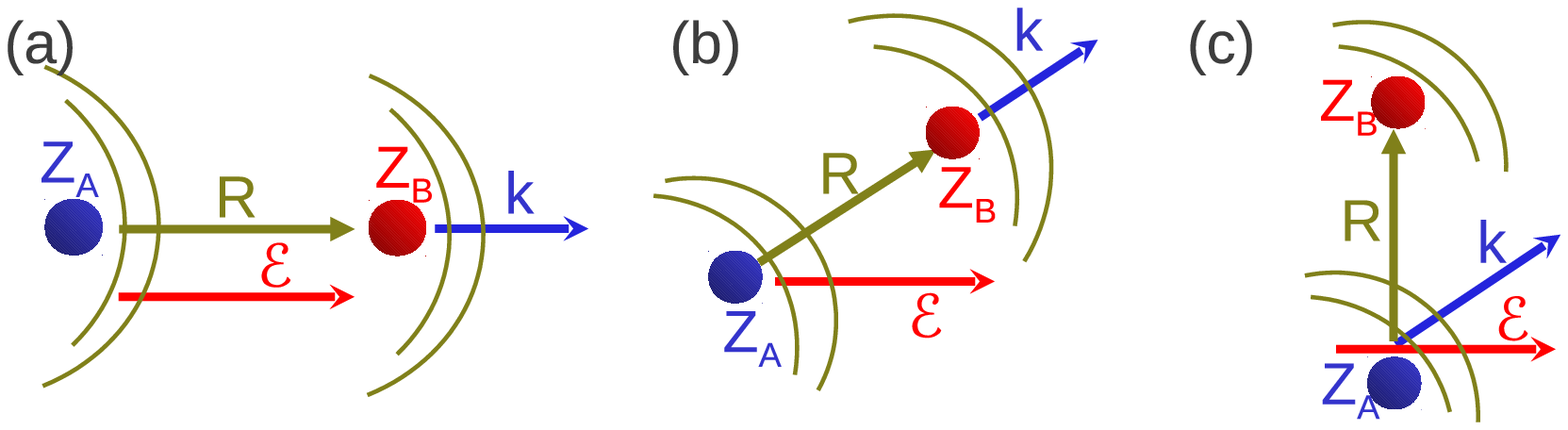}\\
\includegraphics[angle=0, width=\imwidth]{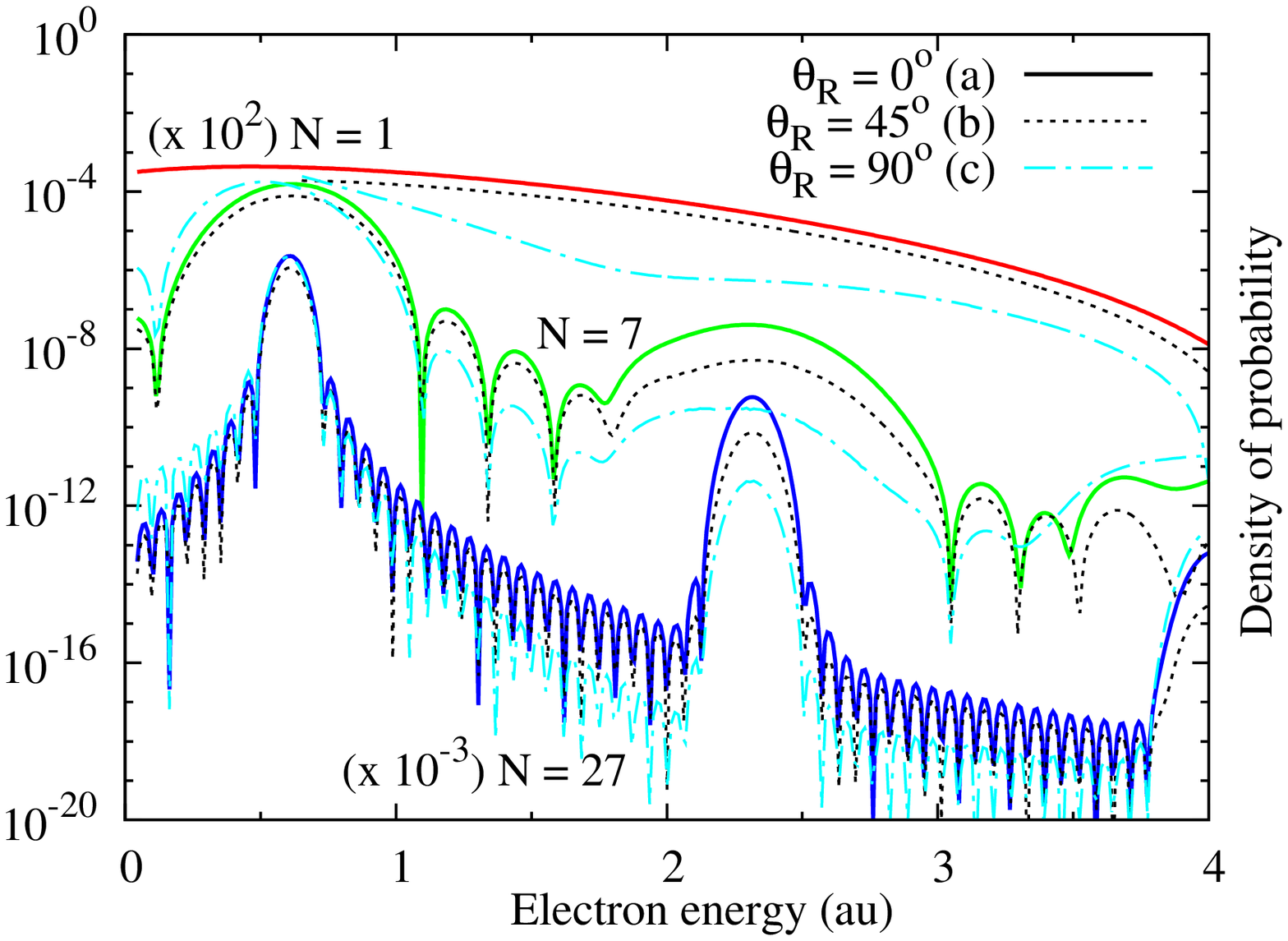}
\caption[]{(colour online) CVA results for $\mathrm{H}_{2}^{+}$ ionisation spectra for the three geometrical arrangement: (a) forward configuration in solid line, (b) $\bi{R}\parallel \bi{k}$ at $45^o$ of $\hat{\varepsilon}$ in dotted line and (c) $\bi{R}\perp  \hat{ \varepsilon}$ and $\bi{k}$ at $45^o$ in dashed dotted line. The laser parameters are the same than in the bottom panel of~\fref{F05w085y171H2}. }
\label{F05w171H2c}
\end{figure}

In \fref{F05w085y171H2} we present the ionisation probability differential in energy and angle, for forward configuration, i.e: the emission direction is parallel to the internuclear axis and both are parallel to the polarisation vector (see \fref{F05w171H2c}(a)). As before, there is good agreement between the CVA (using multipolar expansion up to order 3) and DipA calculations for all the energy-range investigated. The full calculations on CVA are indistinguishable from the third order expansion.

Overall, the spectra for molecular ionisation are similar to the ones presented for atoms (\fref{F05w085Atom} and \ref{F05w171Atom}), but there are some notable differences:
\begin{enumerate}\renewcommand{\labelenumi}{(\arabic{enumi})}
\item The spectra are energy-shifted due to different ionisation potentials of the targets. 
The computed $1s\sigma$ electronic energy is $-1.1\mathrm{~a.u.} +1/R = -16.3$~eV. Since we are considering fixed nuclei, the term $1/R$ is also present in the energy of the final continuum state. Then, the net shift in the spectra respect the atomic case is $-0.6$~a.u.~as it can be seen in the position of ATI peaks in the figures. 

\item The molecular spectra are shown for forward-direction emission, while in the atomic case we have presented the probabilities integrated in emission direction.  The principal difference is that the deep minima in the forward condition, similar to the atomic spectra for fixed direction $\bi{k} \parallel \hat{\varepsilon}$ (not shown), are softened by integration on emission angles.

\item The agreement between CVA and DipA calculations is better in the atomic case. This fact can be understood since the $1s\sigma$ orbital is more extended in space than the atomic $1s$. Then, for the same laser pulse, the approximation $\bi{A}\cdot \bi{r} \ll 1$ is not satisfied in all range of the integral $g(t)$.
\end{enumerate}
To analyse the emission in other directions we present in \fref{F05w171H2c} the ionisation probabilities in three geometrical configurations: (a) in the forward direction, where the three vectors are parallel, i.e.~$\theta_R=\theta_k=0$, (b) with the electronic momentum parallel to the internuclear axis and both are at $45^{\circ}$ respect to the polarisation vector, i.e.~$\theta_R=\theta_k=\pi/4$, and (c) where the molecule is perpendicular to $\hat{\varepsilon}$,  $\theta_R=90^{\circ}$ and $\theta_k=45^{\circ}$.
In all these cases the spectra are well reproduced by DipA (not shown). In particular, the arrangements (b) and (c) have the same contribution of $\hat{\varepsilon}\cdot\bi{k}$. Then, the factor $\bi{M}^{(1)}$ is exactly the same in both cases, and all dependence of the spectra with the molecular orientation arises from the one-photon ionisation transition-matrix.

\begin{figure}
\centering
\includegraphics[angle=0, width=\imwidth]{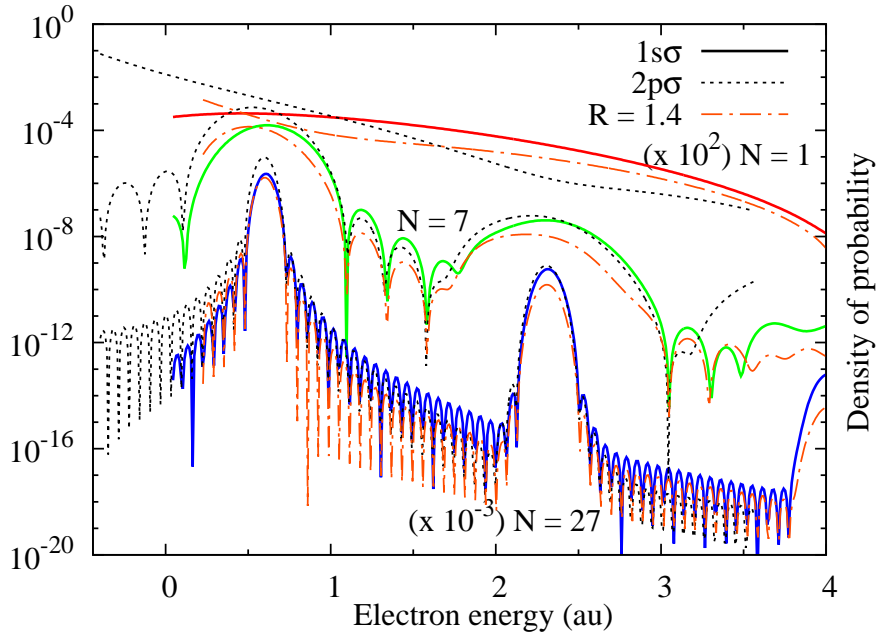}
\caption[]{(colour online) CVA ionisation spectra for $\mathrm{H}_{2}^{+}$ in forward condition from (a) $1s\sigma$ and $R=2$~a.u. in solid line, (b) $2p\sigma$ and $R =2$~a.u. in dotted line, and (c) $1s\sigma$ and $R= 1.4$~a.u. in dashed-dotted line. The laser parameters are the same of \fref{F05w085y171H2}. Curves (b) and (c) are shifted in energy.}
\label{F05w171H2b}
\end{figure}

Similar conclusions can be deduced from the analysis of \fref{F05w171H2b}, where we compare the CVA results for ionisation from $1s\sigma$, $2p\sigma$, in both cases with $R=2$~a.u.~and from $1s\sigma$ at $R= 1.4$~a.u. This value is the equilibrium internuclear distance of the ground-state for the $\mathrm{H}_2$ molecule. As before, the differences are originated in the factor $\bi{L}^{(1)}$.

As concluded from the analysis on molecular ionisation, the dependence on the emission direction, molecular orientation, and molecular structure are derived from the one-photon transition matrix properties.
These results are consistent with reports on previous works of laser-pulse ionisation of $\mathrm{H}_2$ and $\mathrm{H}_2^+$ where several authors have observed coincidences with the one-photon case \cite{Yuan2011PRAp43418,Fernand2009JPBp5602,Fernand2009PRAp63406,Dellapi2007PRAp032710,Selsto2006PRAp033407}. Our present work formalizes this relation and shows that the possibilities of finding ``new molecular effects'' due to the interaction with laser pulses should be investigated outside the range where DipA reproduces well the spectra.

%%%%%%%%%%%%%%%%%%%%%%%%%%%%%%%%%%%%%%%%%%%
\section{Conclusions}

Based on the Coulomb-Volkov approximation (CVA) we have proposed and developed a multipolar expansion for laser-induced ionisation of atoms and molecules. 
In this approach, not only the first but each term is written as the product of two factors. 
One of them comprises the effect of the laser pulse while the second contains all the information of the target structure.

Since the multipolar expansion is based on the Coulomb-Volkov approximation, in section \ref{ApplyCVA} we have explicitly discussed the regime of validity of CVA and analysed additional restrictions for the present expansion. The first order approximation, namely DipA, 
can be applied in the regime of validity of CVA, while simultaneously being in the multiphotonic regime ($\Gamma \gg 1$). 
The main feature of DipA is that it expresses a directly measurable magnitude, the ionisation spectra, as a product of two factors. In this first order, one of them is the well-known one-photon ionisation probability.

Furthermore, although DipA is a first-order approach it accounts for multiphotonic processes and reproduces exactly the ionisation spectra in the region where the one-photon process is dominant (first ATI peak), and therefore also describes accurately the total rate. However it underestimates noticeably the high-order ATI peaks, whose description require second and third-order terms.

We have investigated the accuracy of the first few terms of the multipolar expansion for the ionisation of atoms and molecular-ions of hydrogen induced by short-pulse lasers. 
Differential and total ionisation probabilities were calculated using the Coulomb-Volkov (CVA) and the multipolar expansion, for hydrogen atoms and molecular-ions in the ground-state and some excited states employing exact initial and final target wavefunctions. We were able to analyse the resulting spectra in terms of the separated contributions of laser and target, as expressed in the developed approach.

We have proposed some applications of DipA, covering several aspects of laser-matter interaction, such as the interpretation of the features of the ATI spectra into separated contributions from laser and target structure. Furthermore, the proposed approximations to first, second and third order are concrete examples of computationally low-cost alternative methods to calculate differential and total cross-sections in complex atomic and molecular targets.

%In all cases we discussed the limits of validity and range of applicability of the approximations.

%%%%%%%%%%%%%%%%%%%%%%%%%%%%%%%%%%%%%%%%%%%5

\section{Acknowledgements}

This work was partially supported by the Consejo Nacional de Investigaciones Cient\'{\i}ficas y T\'{e}cnicas (Grant PIP 112-2009-0100166), Universidad Nacional de Cuyo (Grant 06/C340), Argentina and from the EU Seventh Framework Programme under grant agreement PIRSES-GA-2010-269243. 

%%%%%  PARA bibtex  %%%%%%%%%%%%%%%%%%%%%%%%%%%%%%%%%%%%%%%
% \bibliography{manuslaser}
% 
%%%%% Contenido de manuslaser.bbl %%%%%%%%%%%%%%%%%%%%%%%%%

\end{document}